\documentclass{svmult}

\usepackage{makeidx}         
\usepackage{graphicx}        
\usepackage{multicol}        
\usepackage[bottom]{footmisc}

\makeindex             

\begin{document}

\title*{The Rich {\em Are} Different !\\
{\Large Pareto Law from asymmetric interactions
in asset exchange models}}
\titlerunning{The Rich Are Different !} 
\author{Sitabhra Sinha}
\institute{The Institute of Mathematical Sciences, C. I. T. Campus,
Taramani,\\ 
Chennai - 600 113, India.\\
\texttt{sitabhra@imsc.res.in}}

\maketitle
\begin{quote}
{\em Fitzgerald}: The rich are different from you and me

{\em Hemingway}: Yes, they have more money
\end{quote}
\noindent
It is known that asset exchange models with symmetric interaction between 
agents show either a Gibbs/log-normal distribution of assets among the agents 
or condensation of the entire
wealth in the hands of a single agent, depending upon the rules of exchange. 
Here we explore the effects of introducing asymmetry in the interaction 
between agents with different amounts of wealth (i.e., the rich behave
differently from the poor). This can be implemented 
in several ways: e.g., (1) in the net amount of wealth that is transferred 
from one agent to another during an exchange interaction, or (2) the 
probability of gaining vs. losing a net amount of wealth from an exchange 
interaction. We propose that, in general, the introduction of asymmetry leads 
to Pareto-like power law distribution of wealth.  

\vspace{-0.4cm}
\section{Introduction}
\vspace{-0.1cm}
\label{sec:1}
\begin{quote}
``The history of all hitherto existing society is a history of 
social hierarchy" -- {\em Joseph Persky} \cite{Per92}
\end{quote}
\vspace{-0.1cm}
As is evident from the above quotation, the inequality of wealth (and income)
distribution in society has long been common knowledge. However,
it was not until the 1890s that the nature of this inequality was sought
to be quantitatively established. Vilfredo Pareto collected data about the
distribution of income across several European countries, and stated that,
for the high-income range,
the probability that a given individual has income greater than or equal to
$x$ is $P_> ( x ) \sim x^{-\alpha}$, $\alpha$ being known as the Pareto exponent
\cite{Par97}. Pareto had observed $\alpha$ to vary around 1.5 for the
data available to him and believed $\alpha \simeq 1.5$ to be universal 
(i.e., valid across different societies). 
However, it is now known that $\alpha$ can vary
over a very wide range \cite{Fuj03}; 
furthermore, for the low-income end, the distribution
follows either a log-normal \cite{Sou01} or 
exponential distribution \cite{Dra01}. Similar 
power law tails have been observed for the wealth distribution in 
different societies.
While wealth and income are obviously not independent of each other,
the exact relation between the two is not very clear. While wealth is
analogous to {\em voltage}, being the net value of assets owned at a given 
point of time, income is analogous to {\em current}, as it is the net
flow of wages, dividends, interest payments, etc. over a period of time.
In general, it has been observed that wealth is more unequally distributed 
than income. Therefore, the Pareto exponent for wealth distribution 
is smaller than
that for income distribution.

\begin{figure}[tbp]
\centering
\includegraphics[width=0.48\linewidth,clip]{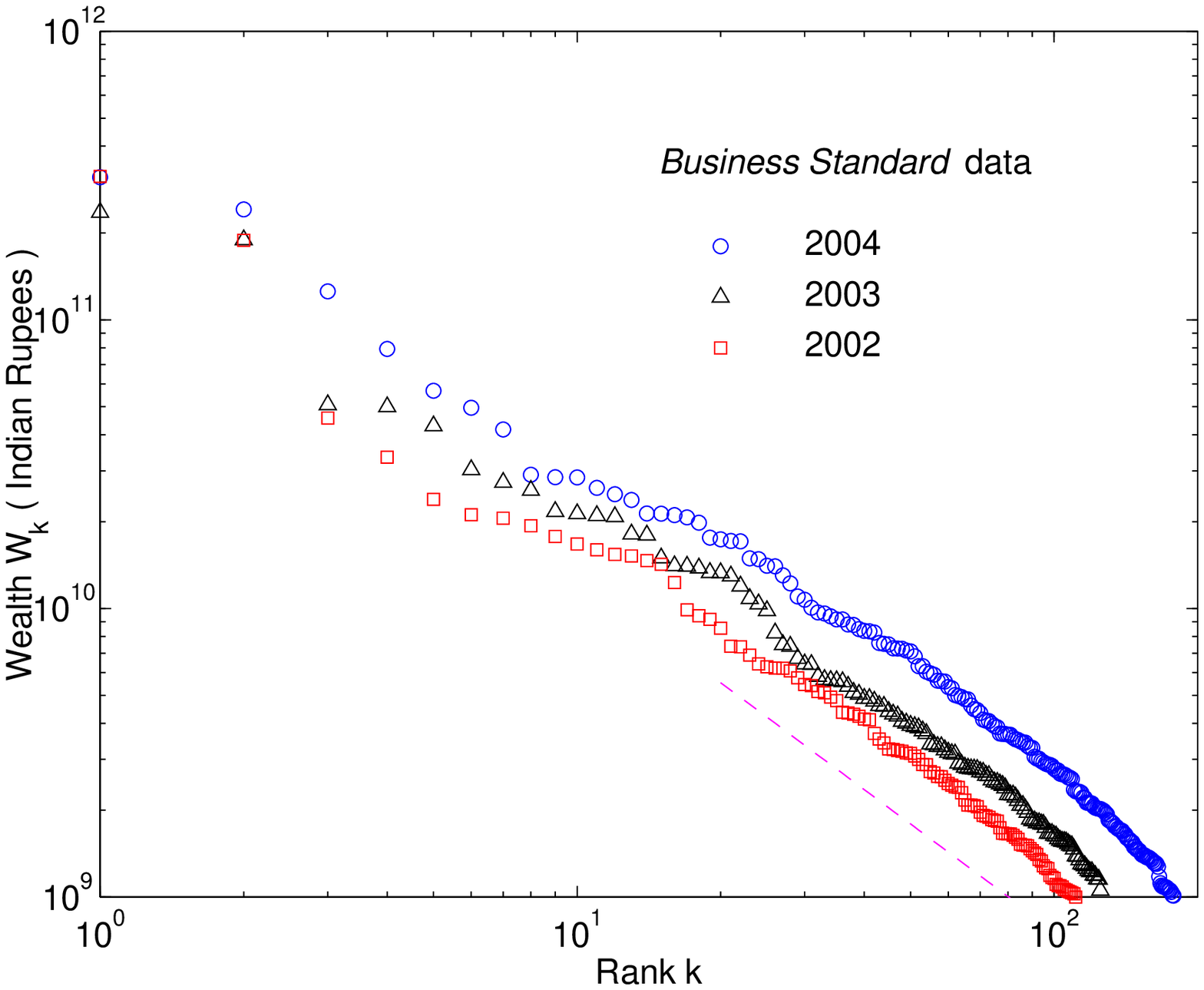}
\includegraphics[width=0.48\linewidth,clip]{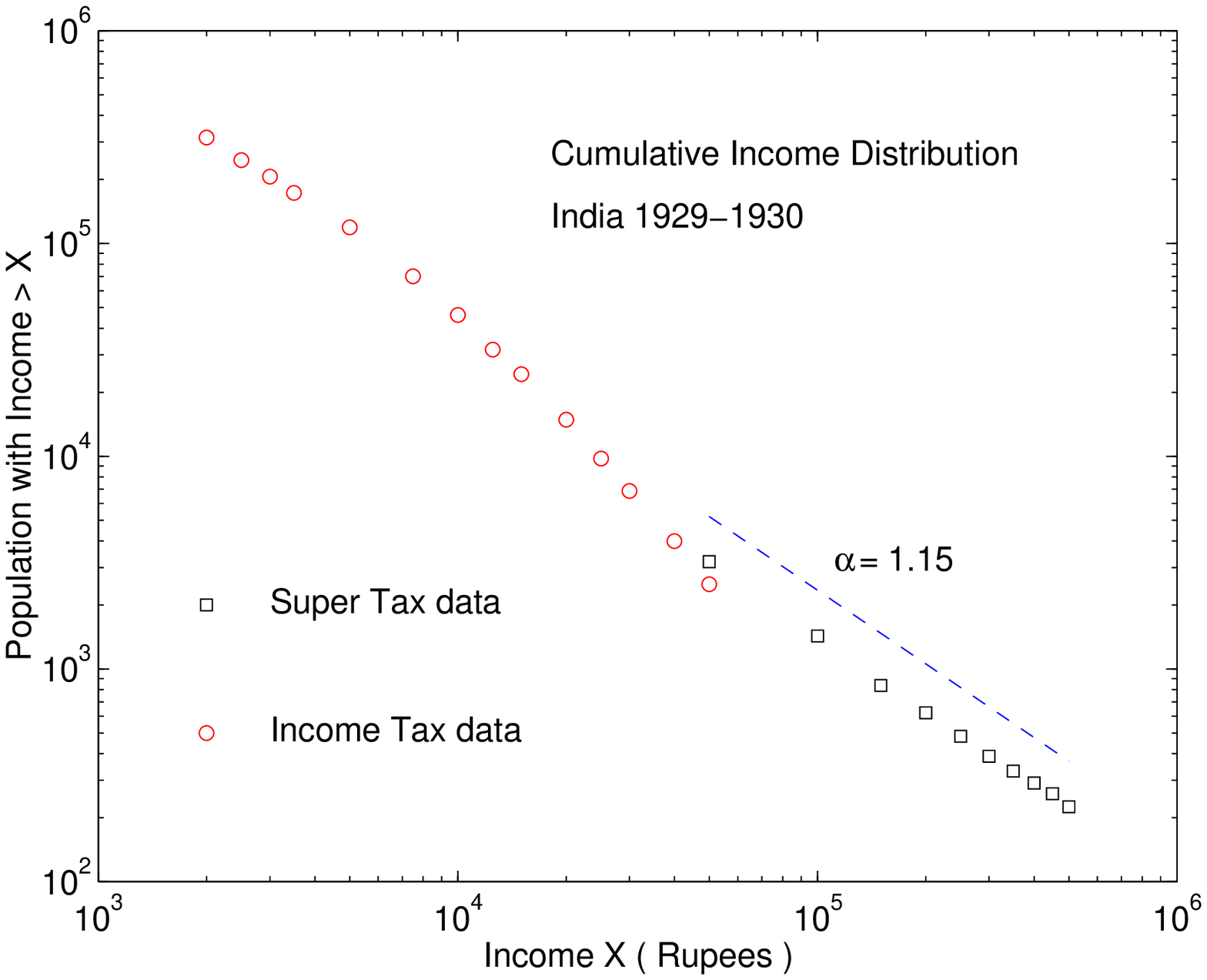}
\caption{Wealth and income distribution in India: (Left) Rank ordered wealth 
distribution during the period 2002-2004 plotted 
on a double-logarithmic scale, showing the wealth of the $k$-th ranked
richest person (or household) in India
against the rank $k$ (with rank 1 corresponding to the
wealthiest person) as per surveys conducted 
by {\em Business Standard} \cite{BusSta}
in Dec 31, 2002 (squares), Aug 31, 2003 (triangles) and Aug 31, 2004 
(circles). The broken line
having a slope of $-1.23$ is shown for visual reference. (Right) Cumulative
income distribution during the period 1929-30 as per information obtained
from Income Tax and Super Tax data given in Ref.~\cite{Shi35}. The plot
has Gibbs/log-normal form at the lower income range, and a power law tail
with Pareto exponent $\alpha \simeq 1.15$ for the highest income range.}
\label{fig:1}       
\vspace{-0.2cm}
\end{figure}
\noindent
Most of the empirical studies on income and wealth distribution have been
done for advanced capitalist economies, such as, Japan and USA.
It is interesting to note that similar distributions can be
observed even for India \cite{Sin05}, which until recently had followed
a planned economy. As income tax and other records about individual holdings
are not publicly available in India, we had to resort to indirect methods.
As explained in detail in Ref. \cite{Sin05}, the
Pareto exponent for the power-law tail of the wealth distribution
was determined from the rank-ordered plot of wealth of the richest
Indians 
[Fig. \ref{fig:1}~(left)].
This procedure yielded 
an average Pareto exponent of $\simeq 1/1.23 = 0.82$. 
A similar exercise carried 
out for the income distribution in the highest income range produced
a Pareto exponent $\alpha \simeq 1.51$. Surprisingly, this is identical
to what Pareto had thought to be the universal value of $\alpha$. 
Comparing this with historical data of income distribution
in India \cite{Shi35}, we again observe the power-law tail although with
a different exponent [Fig. \ref{fig:1}~(right)]. In addition, we note 
that the low-income range has a log-normal or Gibbs form very similar
to what has been observed for advanced capitalist economies \cite{Sou01}.
In the subsequent sections, we will try to reproduce these observed features
of wealth \& income distributions through models belonging to the general
class of asset exchange models.

\vspace{-0.4cm}
\section{Asset exchange models}
\vspace{-0.2cm}
Asset exchange models belong to a class of simple models of a closed economic
system, where the total wealth available for exchange, $W$, and the total
number of agents, $N$, trading among each other, are 
fixed \cite{Isp98,Cha00,Dra00,Sin03,Cha04}. 
Each agent $i$ has some wealth $W_i (t)$ associated with it at
time step $t$. Starting from an arbitrary initial distribution of wealth
($W_i (0)$, $i = 1, 2, 3, \ldots.$), during each time step two randomly chosen
agents $i$ and $j$ exchange wealth, subject to the constraint that the 
combined wealth of the two agents
is conserved by the trade, and that neither of the two has
negative wealth after the trade (i.e., debt is not allowed).
In general, one of the
players will gain and the other player will lose as a result of the trade.
If we consider an arbitrarily chosen pair of agents ($i$, $j$) who trade
at a time step $t$, resulting in a net gain of wealth by agent $i$,
then the change in their wealth as a result of trading is:
$$
W_i (t + 1) = W_i (t) + \Delta W; W_j (t + 1) = W_j (t) - \Delta W,
$$
where, $\Delta W$ is the net wealth exchanged between the two agents.

\begin{figure}[htbp]
\centering
\includegraphics[width=0.48\linewidth,clip]{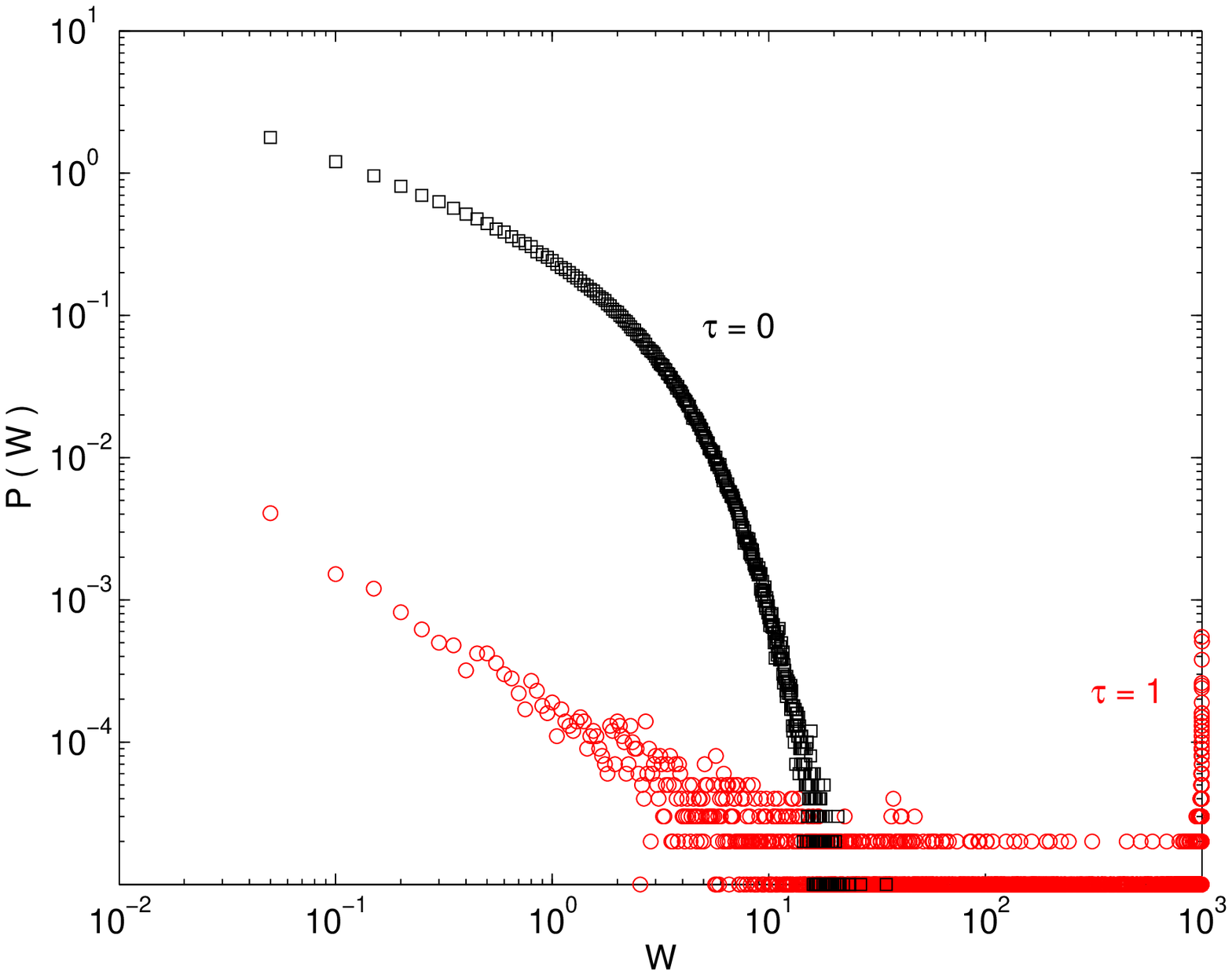}
\includegraphics[width=0.48\linewidth,clip]{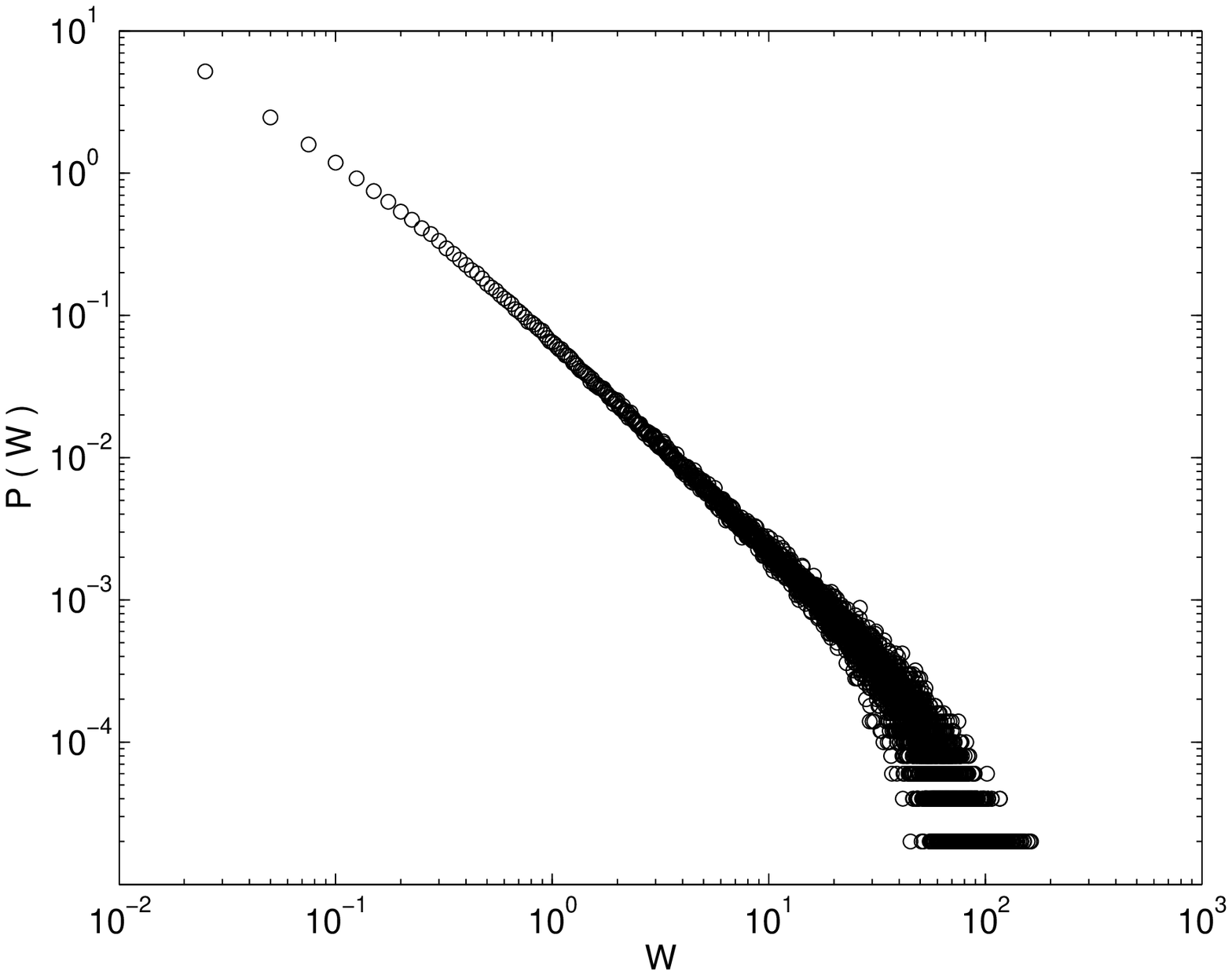}
\caption{(Left) Asymptotic wealth distribution for the random 
exchange model ($\tau = 0$: exponential distribution) and the minimum 
exchange model ($\tau =1$: condensation). (Right) Power law wealth
distribution with exponent $\simeq -1.5$ for the asymmetric exchange model
with $\tau = 0.99$. All figures shown for $N = 1000$,
$t = 1 \times 10^7$ iterations, averaged over 2000 realizations.
}
\label{fig:2}       
\vspace{-0.2cm}
\end{figure}

\noindent
Different exchange models are defined based on how $\Delta W$ is related
to $[ W_i (t), W_j (t) ]$. For the {\em random exchange} model, the wealth 
exchanged is a random fraction of
the combined wealth [$W_i (t) + W_j (t)$], while for the {\em minimum
exchange} model, it is a random fraction of the wealth of the poorer
agent, i.e., $ min [ W_i (t), W_j (t) ]$]. The asymptotic distribution for the
former is exponential, while the latter shows a condensation of the entire
wealth $W$ into the hands of a single agent [Fig.~\ref{fig:2}~(left)]. 
Neither of these reflect the
empirically observed distributions of wealth in society, discussed in the
previous section.

\noindent
Introducing savings propensity in the exchange mechanism, whereby agents
don't put at stake (and are therefore liable to lose) their entire
wealth, but put in reserve a fraction of their current holdings, does
not significantly change the steady state distribution \cite{Cha00}. 
By increasing the savings 
fraction (i.e., the fraction of wealth of an agent that is not being put at
stake during a trade), one observes that the steady-state distribution becomes 
non-monotonic, although the tail still decays exponentially.
However, randomly assigning different savings fractions (between [0,1]) to
agents lead to a power-law tail in the asymptotic distribution \cite{Cha04}.

\noindent
This result raises the question of whether it is the differential ability
of agents to save that gives rise to the Pareto distribution.
Or, turning the question around, we may ask whether the rich save more. 
This question 
has been the
subject of much controversy, but recent work seems to have answered this
in the affirmative \cite{Dyn04}. As mentioned in a leading economics
textbook, 
savings is the greatest luxury of all \cite{Sam01} and the amount of savings
in a household rises with income. In terms of the asset exchange models,
one can say that an agent with more wealth is more likely to save (or saves
a higher fraction of its wealth). Implementing this principle appropriately
in the exchange rules, one arrives at the {\em asymmetric exchange} model.
\vspace{-0.4cm}
\section{Asymmetric exchange model}
\vspace{-0.2cm}
\noindent
The model is defined by the following exchange rules specifying the 
change in wealth, $W_A (t+1) - W_A (t)$, of agent $A$ who 
wins a net amount of wealth after trading with agent $B$
[$W_B (t+1) - W_B (t) = W_A (t) - W_A (t+1)$]: 
$$
W_A (t+1) = W_A (t) + \epsilon ( 1 - \tau [1 - \frac{W_A (t)}{W_B (t)}]) 
W_B (t), ~{\rm if}~ W_A (t) \leq W_B (t),
$$
$$
= W_A (t) + \epsilon W_B (t), {\rm otherwise},
$$
where $\epsilon$ is a random number between 0 and 1, specifying the
fraction of wealth that has been exchanged.
For $\tau = 0$, this is the random exchange model, while for $\tau = 1$,
it is identical to the minimum exchange model [Fig. \ref{fig:2} (left)].
In the general case, $0 < \tau < 1$, the relation between the agents trading 
with each other
is asymmetric, the richer agent having more power to dictate terms of trade
than the poorer agent. The parameter $\tau$ ({\em thrift}) 
measures the degree to which the richer agent is able to use this power.

\noindent
As $\tau$ is increased from 0 to 1, the asymptotic distribution of wealth
is observed to change from exponential to a condensate (all wealth belonging
to a single agent). However, at the transition between these two very
different types of distribution ($\tau \rightarrow 1$) one observes 
a power-law distribution ! As seen in Fig.~\ref{fig:2}~(right), the
power-law extends for almost the entire range of wealth and has a Pareto
exponent $\simeq 0.5$. This is possibly the simplest asset exchange model that 
can give rise to a power-law distribution. Note that, unlike other models
\cite{Cha04},
here one does not need to assume the distribution of a parameter
among agents.

\begin{figure}[tbp]
\centering
\includegraphics[width=0.48\linewidth,clip]{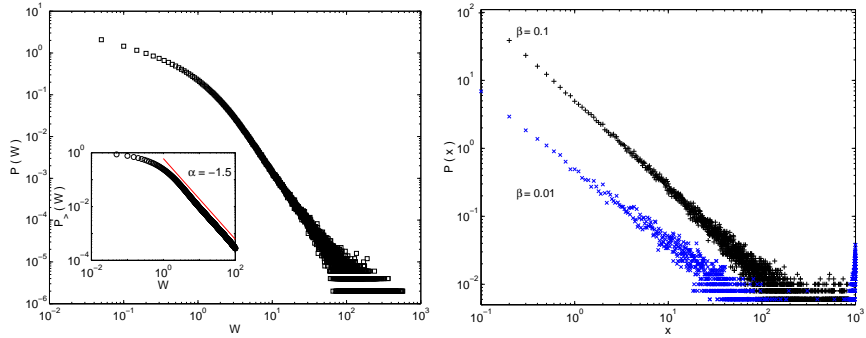}
\includegraphics[width=0.48\linewidth,clip]{ss_uasp03_fig7b.eps}
\caption{(Left) Asymptotic wealth distribution (inset shows the cumulative
distribution) with a power-law tail having 
Pareto exponent $\alpha \simeq 1.5$, for the asymmetric exchange model 
with $\tau$ distributed uniformly over the unit interval $[0, 1]$
among $N$ agents ($N = 1000$, $t = 1 \times 10^7$ iterations, averaged 
over $10^4$ realizations). (Right) Asymptotic wealth distribution 
for model having asymmetric winning probability with $\beta = 0.1$ [pluses]
(slope of the power-law curve is $1.30 \pm 0.05$)
and $\beta = 0.01$ [crosses] (slope of the power-law curve 
is $1.27 \pm 0.05$). ($N = 1000$, $t = 1.5 \times 10^7$ iterations,
averaged over $5000$ realizations).
}\label{fig:3}
\vspace{-0.2cm}
\end{figure}

\noindent
However, the Pareto exponent for this model is quite small compared to 
those empirically observed in real economies. This situation is remedied 
if instead of considering a fixed value of $\tau$ for all agents, we 
consider the heterogeneous case where $\tau$ is distributed randomly among 
agents according to a quenched distribution. For an uniform distribution
of $\tau$, 
the steady-state distribution of wealth has a power-law tail
with $\alpha = 1.5$ [Fig.~\ref{fig:3}~(left)],
which is the value predicted by Pareto, while at the region
corresponding to low wealth, the distribution is exponential. 
By changing the nature of the random distribution, one observes power-law
tails with different exponents. For example, for $P ( \tau ) \sim \tau$,
the resulting distribution has a Pareto exponent $\alpha \sim 1.3$, while
for $P ( \tau ) \sim \tau^{-2/3}$, one obtains $\alpha \sim 2.1$. A
non-monotonic U-shaped distribution of $\tau$ yields $\alpha \sim 0.73$.
However, the fact that even with these extremely different distributions
of $\tau$ one always obtains a power-law tail for the wealth distribution,
underlines the robustness of our result.

\vspace{-0.4cm}
\section{Asymmetric Winning Probability Model}
\vspace{-0.2cm}
Asymmetry in the interaction between agents (as a function of their wealth)
can also be introduced through the probability that an agent will gain
net wealth from an exchange. Consider a variant of the minimum exchange 
model
where the probability that agent $A$ (wealth $W_A$) will win a net amount
in an exchange with $B$ (wealth $W_B$) is
$$
p (A|A,B) = \frac{1}{1 + exp ( \beta [\frac{W_A (t)}{W_B (t)} - 1])},
$$
where $\frac{1}{\beta}$ is the indifference to relative wealth 
(for details see Ref.~\cite{Sin03}).
For $\beta$ = 0, i.e., $p (A|A,B) = \frac{1}{2}$,
the minimum exchange model is retrieved, where, in the steady state,
the entire wealth belongs
to a single agent (condensation). However, for 
a finite value of
$\beta$, the poorer agent has a higher probability of
winning. For large $\beta$, the asymptotic distribution is 
exponential, similar to the random exchange model.
At the transition between these two very different types of distributions
(condensate and exponential) we observe a power-law distribution
of wealth [Fig. \ref{fig:3} (right)].
\vspace{-0.4cm}
\section{Discussion}
\vspace{-0.2cm}
The two models discussed here for generating Pareto-like distribution of 
wealth 
are both instances of 
the ``Rich Are Different'' principle, implemented in the formalism
of asset exchange models. It is interesting to note that other
recently proposed models for generating Pareto law also use this 
principle, whether this is in terms of kinetic theory as in the
present paper \cite{Din03,Igl04} or in a network context
\cite{Ast04,Bha05}. This leads us to conclude that
asymmetry in agent-agent interactions is a crucial feature of models
for generating distributions having power-law tails.

\noindent
To conclude, we have presented two models illustrating the general principle
of how Pareto-like distribution of wealth (as observed in empirical observations
in society) can be reproduced by implementing asymmetric 
interactions between agents in asset exchange models. In the models presented
here the asymmetry is based on wealth of agents, with the rich agents 
behaving differently from the poor, either in terms of net wealth changing
hands, or the probability of gaining net wealth
out of a trade. One of the models is possibly the simplest asset exchange
model that gives a power-law distribution. The results are also very robust,
the power law being observed for a wide variety of parameter distributions.
The different values of $\alpha$ obtained for different
parameter distributions is a possible explanation of why different
Pareto exponents have been measured in different societies, as well as in the
same society at different times.

\vspace{-0.5cm}
%
%
%
%

%
%



\printindex
\end{document}